%% file: main.tex
\documentclass[twocolumn, switch]{article}
\usepackage[top=0.8in, bottom=0.8in, left=0.75in, right=0.75in]{geometry}
\usepackage{amsmath, amsthm, amssymb, amsfonts}

\usepackage{etoolbox}

\AtBeginEnvironment{equation}{\small}
\AtBeginEnvironment{align}{\small}
\AtBeginEnvironment{gather}{\small}

\usepackage[numbers,square]{natbib}
\usepackage[utf8]{inputenc}	
\usepackage[T1]{fontenc}	
\usepackage{xcolor}		
\usepackage[hyphens]{url}
\usepackage[colorlinks = true,
            linkcolor = purple,
            urlcolor  = blue,
            citecolor = cyan,
            anchorcolor = black]{hyperref}	
\usepackage{booktabs} 		
\usepackage{nicefrac}		
\usepackage{microtype}		
\usepackage{lineno}		
\usepackage{float}			
\usepackage{tcolorbox}

\usepackage{newfloat}
\DeclareFloatingEnvironment[name={Supplementary Figure}]{suppfigure}
\usepackage{sidecap}
\sidecaptionvpos{figure}{c}

\usepackage{titlesec}
\titlespacing\section{0pt}{10pt plus 3pt minus 3pt}{1pt plus 1pt minus 1pt}
\titlespacing\subsection{0pt}{10pt plus 3pt minus 3pt}{1pt plus 1pt minus 1pt}
\titlespacing\subsubsection{0pt}{8pt plus 3pt minus 3pt}{1pt plus 1pt minus 1pt}
\titlespacing\section{0pt}{12pt plus 3pt minus 3pt}{1pt plus 1pt minus 1pt}
\titlespacing\subsection{0pt}{10pt plus 3pt minus 3pt}{1pt plus 1pt minus 1pt}
\titlespacing\subsubsection{0pt}{8pt plus 3pt minus 3pt}{1pt plus 1pt minus 1pt}

\usepackage{siunitx}
\usepackage{subcaption}
\usepackage{optidef}
\usepackage{multirow}

\usepackage{enumitem}
\setlist{nosep}  

\usepackage{tikz,hyperref}

\definecolor{lime}{HTML}{A6CE39}
\DeclareRobustCommand{\orcidicon}{
	\begin{tikzpicture}
	\draw[lime, fill=lime] (0,0)
	circle [radius=0.16]
	node[white] {{\fontfamily{qag}\selectfont \tiny ID}};
	\draw[white, fill=white] (-0.0625,0.095)
	circle [radius=0.007];
	\end{tikzpicture}
	\hspace{-2mm}
}
\foreach \x in {A, ..., Z}{\expandafter\xdef\csname orcid\x\endcsname{\noexpand\href{https://orcid.org/\csname orcidauthor\x\endcsname}
			{\noexpand\orcidicon}}
}

\title{CEO-DC:~Driving~Decarbonization in~HPC~Data~Centers~with~Actionable~Insights}
\date{}

\usepackage{authblk}

\newcommand{\equalcontrib}{\textsuperscript{\dag}}

\author[1*]{Rubén~Rodríguez~Álvarez\equalcontrib\orcidA{}}
\author[1]{Denisa-Andreea~Constantinescu\equalcontrib\orcidB{}}
\author[2]{Miguel~Peón-Quirós\orcidC{}}
\author[1]{David~Atienza\orcidD{}}

\affil[1]{Embedded Systems Laboratory, École Polytechnique Fédérale de Lausanne (EPFL), Switzerland}
\affil[2]{EcoCloud Center, École Polytechnique Fédérale de Lausanne (EPFL), Switzerland}
\affil[*]{\tt ruben.rodriguezalvarez@epfl.ch}

\usepackage{titlesec}

\titleformat{\section}
  {\large\bfseries\raggedright\hyphenpenalty=10000}
  {\thesection}{1em}{}

\titleformat{\subsection}
  {\normalsize\bfseries\raggedright\hyphenpenalty=10000}
  {\thesubsection}{1em}{}

\titleformat{\subsubsection}[runin]
  {\normalsize\bfseries} 
  {\thesubsubsection.}
  {1em}
  {}
  [.]

\begin{document}

\newcommand{\ecodc}{CEO-DC }
\newcommand{\keywords}[1]{%
  \noindent\textbf{Keywords: } #1
  \vspace{0.5em}
}

\twocolumn[
  \begin{@twocolumnfalse}

\maketitle

\vspace{-3em}
\begin{abstract}

The rapid growth of data centers is increasing energy demand and widening the carbon gap in the ICT sector, as fossil fuels still dominate global energy production. Addressing this challenge requires collaboration across research, policy, and industry to rethink how computing infrastructures are designed and scaled sustainably. This work addresses central trade-offs in procurement decisions that affect carbon emissions, economic costs, and scaling of compute resources. We present these factors in a holistic decision-making framework for Carbon and Economy Optimization in Data Centers (CEO-DC). CEO-DC introduces new carbon and price metrics that enable DC managers, platform designers, and policymakers to make informed decisions. Applying CEO-DC to current trends in AI and HPC reveals that, in \SI{72}{\percent} of the cases, platform improvements lag behind demand growth. Moreover, prioritizing energy efficiency over latency can reduce the economic appeal of sustainable designs. Our analysis shows that in many countries with electricity with medium to high carbon intensity, replacing platforms older than four years could reduce their projected emissions by at least \SI{75}{\percent}. However, current carbon incentives worldwide remain insufficient to steer data center procurement strategies toward sustainable goals. In summary, our findings underscore the need for a shift in hardware design and faster grid decarbonization to ensure sustainability and technological viability.

\end{abstract}
\vspace{0.35cm}

\keywords{
Data Centers, Sustainability, Carbon Gap, Carbon Emissions, Incentives, Design Space Exploration, Procurement Strategies, Energy Efficiency, AI, HPC
}

  \end{@twocolumnfalse}
]

\NoHyper
\begingroup
  \renewcommand\thefootnote{\dag}
  \footnotetext{Both authors contributed equally to this work.}%
\endgroup
\endNoHyper

\tcbset{
  colback=gray!10,
  colframe=gray!60!black,
  boxrule=0.3pt,        
  arc=2pt,              
  left=1mm,             
  right=1mm,            
  boxsep=1pt,           
  coltitle=white,
  fonttitle=\bfseries,
  before skip=3pt,
  after skip=3pt,
  parbox=false,
  before upper=\setlength{\parindent}{0pt},
}


\section{Introduction}

\begin{figure}[ht]
    \centering
    \includegraphics[width=0.9\linewidth]{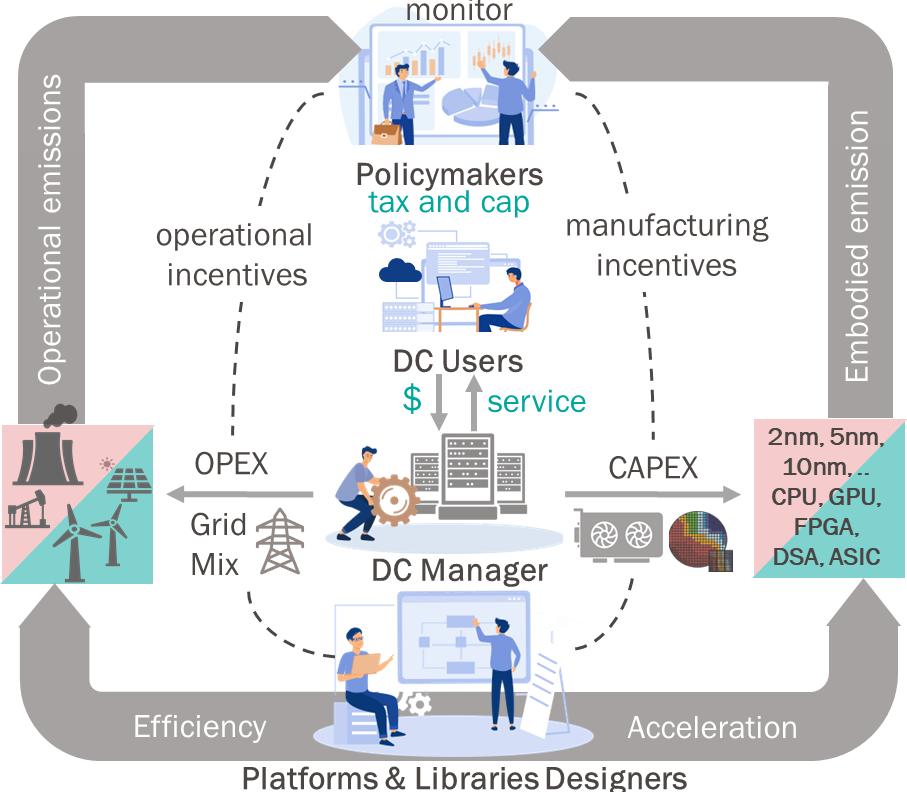}
    \caption{
     Key stakeholders for sustainable DC scaling.
    }
    \label{fig:carbon-flow}
\end{figure}

Data centers (DCs) are the fastest growing electricity consumers in the Information and Communication Technology (ICT) sector \cite{Masanet_2020, Jones_2018, CSO2023DCMEC, IEAelec2024}. The use of electricity by Google, Meta and Microsoft has more than doubled between 2017 and 2021, with AI as the main driver \cite{wu_meta_2024}, while the International Energy Agency (IEA) projects that global DC energy consumption could double from \SI{460}{TWh}  in 2022 to exceed \SI{1000}{TWh} in 2026 \cite{IEAelec2024}.
Unconstrained growth will drive a sharp rise in greenhouse gas emissions in the next five years \cite{Freitag_2021, IEAelec2024, kairos2025google, eeckhout2025sustainability, aslan_2025}. 
These trends underscore the growing challenge of closing the emissions gap in DCs, calling for urgent decarbonization efforts to meet the sustainability goals of the Paris Agreement.

Addressing this gap requires holistic approaches that capture the complex interdependencies throughout the lifecycle of the compute system and identify the optimal Pareto frontiers for carbon, performance, and cost efficiency~\cite{Eliam_2024_vision_specialization_dc}. Developing fair and accurate sustainability models for DCs requires a rigorous integration of computing requirements with economic realities, while explicitly accounting for future demand growth to prevent overestimating the long-term sustainability benefits of efficiency gains~\cite{luccioni_jevons_effect_2025}.
The challenge lies in aligning investment strategies to design, procure, and manage computing platforms that achieve climate-responsible growth while maintaining economic viability. However, existing research provides limited guidance for DC operators striving to make decisions that are simultaneously cost-effective and environmentally sustainable\footnote{For brevity, we use ``sustainable'' to refer to ``environmentally sustainable.''}~\cite{Gandhi2023_metrics_sustainability_dc}. We respond to the call to ``holistically address the growing environmental footprint of computing''~\cite{eeckhout2025sustainability}.

We identify the key DC stakeholders and their interactions, as illustrated in Fig.~\ref{fig:carbon-flow}. \textit{DC managers} meet computing demand from \textit{DC users} by operating facilities and expanding capacity with new platforms. \textit{Platform and library designers} influence the carbon and price efficiency of emerging platforms through their design choices. \textit{Policymakers} monitor and regulate emissions.
We study how these stakeholders can balance sustainability and profitability through six critical questions or trade-offs:
\begin{itemize}
    \item \textbf{Q.1} Do OPEX savings justify CAPEX investments in emissions and costs?
     \item \textbf{Q.2} Is environmental sustainability aligned with economic viability?
     \item \textbf{Q.3} To what extent can a DC expand its compute capacity while still meeting sustainability goals?
    \item \textbf{Q.4} What level of incentive is required to align sustainable decisions with economic drivers?
     \item \textbf{Q.5} How should performance acceleration and energy efficiency be balanced to achieve sustainability and viability?
    \item \textbf{Q.6} When should specialization be favored over flexibility from a sustainability and economic point of view?
\end{itemize}

We present the CEO-DC framework structured around these key trade-offs central to sustainable computing and propose new metrics to assess them. We evaluated this framework in an HPC AI case study, obtaining actionable insights for each stakeholder group. These insights provide a concrete basis to align economic objectives with long-term sustainability goals in the DC industry.

\subsection{Key Contributions}

We propose a novel decision-making framework for Carbon and Economy Optimization in Data Centers (CEO-DC) to address the limitations of current strategies and support a coordinated effort toward sustainable DC growth, integrating the following scientific contributions:
\begin{enumerate}
    \item A \textbf{carbon-economy\footnote{Through this paper, we use the expression ``carbon-economy'' to denote the trade-off between carbon sustainability and economic viability (not to be confused with economic models based on carbon trading).} model} that captures six critical trade-offs throughout the DC lifecycle regarding sustainability, computing growth, and economic interests.
    \item \textbf{Carbon and price efficiency metrics} that evaluate the trade-offs of the model and allow stakeholders to make informed decisions.
    \item \textbf{Real-world case study} on \textit{AI-driven HPC DCs in leading countries}, highlighting the policies and incentives required to meet sustainability goals.
    \item \textbf{Actionable insights} for \textit{stakeholders}, including (i) design space exploration methods for platform designers, (ii) upgrade planning strategies for DC managers, and (iii) quantified carbon incentives and policy levers for policymakers.
\end{enumerate}

The remainder of the paper is organized as follows: Sect.~\ref{sec:background} reviews related work and its key limitations. Sect.~\ref{sec:methodology} presents our carbon-economy framework to quantify and close the carbon gap. Sec.~\ref{sec:results} analyzes results from applying CEO-DC to an AI demand growth case study. Sect.~\ref{sec:discussion} discusses the benefits, limitations, and future directions. Finally, Sect.~\ref{sec:conclusion} concludes the paper.

\section{Current Methods to Quantify Emissions and Optimize DC Procurement} \label{sec:background}

In this section, we review current metrics~\cite{schneider_tpu_lca_2025, Gandhi2023_metrics_sustainability_dc, Gupta_2020} and methods to quantify the total emissions of DC–grade platforms, from a hardware~\cite{ji2024scarif, schneider_tpu_lca_2025, tomlinson2024carbon} and software perspective~\cite{Lannelongue_2021, faiz_llmcarbon_2023}. We also examine carbon reduction strategies for DC procurement~\cite{bansal2023mosaic, Eliam_2024_vision_specialization_dc, Manganelli_2021}, highlighting their limitations and opportunities for improvement.

\subsection{Quantifying Emissions of DC Platforms} \label{sec:sota-quantify}

The life cycle assessment (LCA) ISO 14040:2006~\cite{finkbeiner2006new} has guided the development of many tools for assessing total carbon emissions in ICT services and platforms.
PAIA~\cite{paia_mit} and Boavizta~\cite{Boavizta_2025} are examples of such tools used by Dell, HP, and Lenovo to report server hardware emissions.
Software-based tools such as GreenAlgorithms~\cite{Lannelongue_2021} and CodeCarbon~\cite{schmidt2021codecarbon} automate the estimation of code-specific operational emissions. 
However, embodied emissions are especially difficult to quantify because of limited disclosure from chip and platform manufacturers. Major server-grade vendors such as Intel, NVIDIA, AMD, and Altera rarely report device-specific embodied carbon data~\cite{Intel2023CR, AMD2023CR}. Academic tools attempt to fill this gap: ACT~\cite{Gupta_2023} derives chip-level embodied emissions from TSMC reports~\cite{TSMC2023Sustainability}, while GreenFPGA and ECO-CHIP~\cite{Sudarshan_2023, Sudarshan_2024} explore ASIC vs. FPGA trade-offs. At the server level, SCARIF~\cite{ji2024scarif} can predict the embodied carbon using vendor data. Li et al.~\cite{li_LLM_LCA_2024} extend ACT for server emissions in the AI application domain, while \cite{faiz_llmcarbon_2023} further contributes application-level insights. 

Recent efforts to define actionable sustainability metrics for computing platforms have yielded useful but fragmented tools. Google introduced the Compute Carbon Intensity (CCI) metric (measured in $gCO_2$-eq per $10^{18}$ FLOP) to benchmark carbon efficiency across TPU generations, although it only reflects operational emissions and excludes embodied carbon and economic trade-offs~\cite{schneider_tpu_lca_2025}.  Gandhi et al.~\cite{Gandhi2023_metrics_sustainability_dc} emphasize that metrics not linked to financial impact often struggle to influence decision making and propose a more comprehensive framework with four metrics that account for OPEX and CAPEX emissions, cost, and quality of service.  However, the proposed metrics require fine-grained job-level carbon accounting, which may limit adoption in production settings. 
Gupta et al.~\cite{Gupta_2020, Gupta_2023} tackle this by introducing CAPEX vs OPEX-dominated classifications, emphasizing embodied carbon in hardware lifecycles, but omit modeling workload growth or upgrade incentives. ECO-CHIP~\cite{Sudarshan_2024} extends sustainability modeling to chiplet-based designs, combining area, power, cost, and carbon estimates. However, its scope is restricted to architectural chiplet variants rather than full DC-grade platforms. Despite this progress, most of the prior work isolates individual factors without capturing their dynamics under demand growth.

We address these limitations by explicitly integrating a novel carbon-economy model into the DC planning and platform design process.
Our model incorporates key contextual factors such as power usage effectiveness (PUE),  carbon intensity, and domain-specific demand growth to inform platform selection and design in terms of performance, energy efficiency, emissions, and costs.
Furthermore, we integrate all identified components that contribute to the embodied emissions of computing platforms in academic studies~\cite{Gupta_2023}, databases~\cite{Boavizta_2025}, and industry tools~\cite{IMECNetZero_2025} into a comprehensive model of embodied emissions. We also include server-level components that are typically replaced during platform upgrades, such as memory, host CPUs, and chassis. This model is used in our decision-making framework to evaluate platform replacement scenarios and provide actionable insights to DC stakeholders in Sect.~\ref{sec:methodology}.

\subsection{Planning DC Procurement for Growth} \label{sec:sota-planning}

Many sustainable DC procurement strategies employ multi-objective optimization for economic cost and carbon footprint~\cite{Manganelli_2021, bansal2023mosaic, Figini_2025}. However, current studies like SCARIF~\cite{ji2024scarif} and ACT~\cite{Gupta_2023} generally address independent applications and single-platform or fixed heterogeneous configurations (e.g., CPU + GPU), lacking support for multi-application, multi-platform upgrade contexts. In the context of cloud LLM servers (CPU and GPU), Li et al.~\cite{li_LLM_LCA_2024} propose an asymmetric upgrade strategy based on the differing evolution of training and inference workloads: while CPU improvements do not justify frequent upgrades (suggesting a four-year lifespan), GPU energy efficiency for inference doubles every three years~\cite{zhu2023intelligent}, prompting a recommendation to upgrade GPUs every two years. 
We model and generalize the exploration of \textit{procurement strategies for sustainable growth} and introduce incentive-based mechanisms to support their implementation in real world settings. Furthermore, we broaden the scope of current works to include modern platforms (e.g., GPUs, TPUs, IPUs) and exemplify their use to explore upgrade scenarios for AI and HPC domains across multiple geographic regions. 

Most prior work focuses exclusively on emissions, without evaluating the viability of the proposed procurement strategies. For example, Sudarshan et al.~\cite{Sudarshan_2023} present GreenFPGA, a comparative analysis of carbon emissions between FPGAs and ASICs across application types, production volumes, and hardware lifespans. 
We advance the state-of-the-art by modeling maximum sustainable growth alongside total emissions and the full cost of the procurement lifecycle. CEO-DC provides a rigorous framework to evaluate both the potential for emission reduction and the economic viability of legacy infrastructure upgrades under real-world carbon policies and demand constraints.

\section{CEO-DC: A Decision Framework for Balancing Carbon and Economic Interests} \label{sec:methodology}

\begin{figure}[t!]
    \centering
    \includegraphics[width=\linewidth]{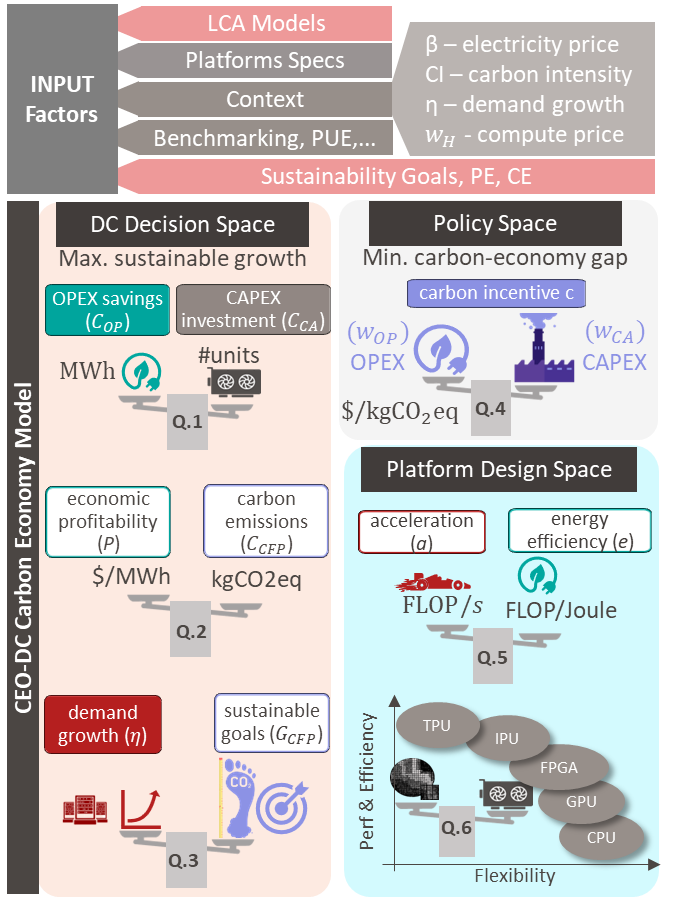}
    \caption{Carbon-economy model.
     Key trade-offs are grouped into the Decision Space (4.1, 4.2, and 4.3), the Policy Space (4.4), and Platform Design Space (4.5 and 4.6).}
    \label{fig:ceo-dc-model}
\end{figure}

This section presents a comprehensive framework to support sustainability-oriented decision-making in DCs, illustrated in Fig.~\ref{fig:ceo-dc-model}.  The framework consists of (i) a model with six core balances to assess sustainability and cost trade-offs for stakeholders in the DC and (ii) four key carbon and price metrics to guide decisions on these trade-offs.
CEO-DC takes as input key factors that influence DC emissions and costs, such as embodied emissions, carbon intensity of electricity generation (CI), and demand growth.

We focus on HPC-dedicated DCs, where predictable workloads enable strategic planning for operations and hardware procurement. We assume that the DC specializes in running specific applications over the hardware lifetime, enabling scheduling flexibility and economies of scale. This is exemplified by NERSC~\cite{nersc}, JSC~\cite{jsc}, CINECA~\cite{cineca}, and ALPS~\cite{alps} HPC DCs are dedicated to advancing the training of foundational AI models, large-scale data analysis in genomics, and scientific discovery in climate, particle physics, and astronomy. Next, we describe and analyze each balance in detail.

\subsection{Input Factors for DC Emissions and Costs} \label{sec:meth-quantify}

We classify the factors that influence the DC procurement decision into: (1) the LCA used when evaluating the embodied footprint of platforms, (2) a benchmark suite representing the applications planned in the DC, (3) contextual variables such as electricity price, CI, compute pricing, and expected demand growth, and (4) sustainability goals.

\subsubsection*{LCA of platforms}

The framework is susceptible to the underlying LCA model used. For this reason, we define a complete formulation to assess the total embodied carbon footprint $C_{\mathrm{CFP}, u,d}$ for a unit $u$ of a platform $d$ as:

\begin{equation}
\label{eq:embodied}
\begin{split}
C_{CFP, u,d} = C_{IC,d} + C_{mem,d} + C_{mb,d} + C_{asm,d} + \\
C_{tp,d} + C_{cool,d} + C_{base,d} /N_{slots}
\end{split}
\end{equation}

This expression incorporates the carbon footprint from
integrated chip (IC) manufacturing and packaging ($C_{IC,d}$), 
memory production ($C_{mem,d}$),
motherboard PCB manufacturing ($C_{mb,d}$),
assembly ($C_{asm,d}$),
transportation to the DC ($C_{tp,d}$),
cooling system impact ($C_{cool,d}$),
and the shared server infrastructure across platforms ($C_{base,d}$) amortized across the number of devices per server ($N_{slots}$).

\subsubsection*{Workload Benchmarking}

An accurate estimation of workload and performance is critical for evaluating alternative hardware. When direct access to platforms is unavailable, relying on vendor peak performance can be misleading due to domain-specific variability. Tools like Intel VTune, Linux \texttt{perf}, and HPCToolkit~\cite{adhianto2010hpctoolkit} help profile energy, performance, and application-level dependencies across platforms. Power usage should include full DC overhead via its PUE.

\subsubsection*{Contextual Data}

It comprises the estimated compute demand growth ($\eta$), the hardware amortization time, the projected electricity price, the CI, and the gross income value \( G_B \)---essential for modeling economic and demand incentives as defined in the sustainable growth balance (Section~\ref{sec:balance3}).

\subsubsection*{Define Sustainability Goals}

A \textit{sustainable goal policy} is the requirement that total emissions $C_{CFP}$ over the system lifetime $T_L$ must not exceed a predefined threshold $G_{CFP}$.
A conservative sustainability goal for DC upgrades can be defined by an \textit{iso-carbon policy}, which aims to achieve total emissions less than or equal to the operations of the legacy system (baseline): $G_{CFP}(T_L) \leq C_{CFP,OP}(M^B)$.

\subsection{Carbon-Economy Model} \label{sec:framework}

When emerging computing platforms offer significant efficiency or performance gains, DC architects are prompted to weigh the benefits of infrastructure upgrades against extending existing hardware lifespans. At the heart of this decision, we have developed a carbon-economy model to analyze trade-offs among operational costs, capital investments, and environmental impact. This model is structured around six key balances following a top-down approach (from low to high level of details), each prompted with a question that serves to guide the decision-making towards sustainability and economic viability. These balances are shown in Fig.~\ref{fig:ceo-dc-model}.

\subsubsection{Balance Between OPEX Savings and CAPEX Investment} \label{sec:balance1}

\textit{Q.1: Do OPEX savings justify CAPEX investments in emissions and costs?}
The \textit{total emissions} and \textit{economic costs} of an upgrade \( C(M^+) \), where \( M^+ \) maps applications to new devices \( D^+ \), are driven by the sum of their respective operational (\( C_{OP} \)) and capital expenditures (\( C_{CA} \)):

\begin{equation}
\label{eq:opex_capex}
C(M^+) = C_{OP}(M^+) + C_{CA}(D^+) \le C_{OP}(M^B)
\end{equation}

This condition informs the strategic decision between upgrading and extending existing infrastructure. An upgrade is preferable when its combined OPEX and CAPEX are lower than the operational cost of the baseline \( C_{OP}(M^B) \). Both cost terms must be comprehensive. OPEX should account for total power consumption of devices and DC PUE, while CAPEX should cover all components replaced during procurement (e.g., server chassis, host CPUs, and memories).

\subsubsection{Balance Between Sustainability and Economic Viability} \label{sec:balance2}

\textit{Q.2: Is environmental sustainability aligned with economic viability?}
Platform upgrade viability and sustainability are evaluated with the baseline workload, where only expenses are accounted for, while DC capacity expansion is treated as a separate decision reflected in the next balance.
An upgrade is classified as \textit{economically viable} when OPEX savings exceed the CAPEX investment (\( V = 1 \) when Eq.~\ref{eq:opex_capex} holds for economic cost). Consequently, if the upgrade reduces net carbon emissions, it is classified as a \textit{sustainable upgrade} (\( S = 1 \) when Eq.~\ref{eq:opex_capex} holds for emissions cost). Economic viability and environmental sustainability often diverge as they are influenced by different and sometimes opposing factors. We define four upgrade outcomes:
\begin{enumerate}
    \item \textit{Naturally incentivized sustainable upgrade} (V=1, S=1)
    \item \textit{Non-incentivized sustainable upgrade} (V=0, S=1)
    \item \textit{Incentivized non-sustainable upgrade} (V=1, S=0)
    \item \textit{Naturally incentivized lifetime extension} (V=0, S=0)
\end{enumerate}

\subsubsection{Balance Between Demand Growth and Sustainability} \label{sec:balance3}

\textit{Q.3: To what extent can a DC expand compute capacity while still meeting sustainability goals?}
This section formalizes the trade-off between scaling computational capacity and meeting sustainability goals. A DC manager is economically incentivized to scale with demand growth. We define the net profit $P$ from a procurement as:
\begin{equation}
\label{eq:P}
\begin{split}
P = \left(G_{B} - C^{\eta=1}_{OP,F}(M^+) - C^{\eta=1}_{CA,F}(D^+)\right) \times \eta + \\
+ (\neg U) \times \left(C^{\eta=1}_{OP,F}(M^+) + C^{\eta=1}_{CA,F}(D^+) - C_{OP,F}(M^B)\right)
\end{split}
\end{equation}

\noindent
where \( G_B \) is the gross income value generated at baseline demand, \( \eta \) represents demand growth, and \( C^{\eta=1}_{OP,F} \), \( C^{\eta=1}_{CA,F} \) are the financial operational and capital costs for the new infrastructure under baseline load (\( \eta=1 \)). The binary variable \( U \in \{0,1\} \) indicates whether legacy hardware is decommissioned/repurposed (\( U=1 \)) or retained (\( U=0 \)). The first term captures the profit from meeting the scaled demand $\eta$ when legacy hardware is replaced. The second term reflects the additional costs (or gains) of maintaining legacy hardware. If keeping the legacy hardware leads to losses in profit (independently of $\eta$), the decision-maker would prefer to upgrade.

Like profit, the cost of the carbon footprint $C_{CFP}$ increases with increasing demand. Its definition is symmetrical to Eq.~\ref{eq:P}, while substituting operational and capital costs for carbon footprint of operations \( C^{\eta=1}_{OP,CFP} \) and manufacturing \( C^{\eta=1}_{CA,CFP}\):

\begin{equation}
\label{eq:CFP_growth}
\begin{split}
C_{CFP} = \left(C^{\eta=1}_{OP,CFP}(M^+) + C^{\eta=1}_{CA,CFP}(D^+)\right) \times \eta + (\neg U) \times\\
 \times \left(C_{OP,CFP}(M^B) - C^{\eta=1}_{CA,CFP}(M^+) - C^{\eta=1}_{CA,CFP}(D^+)\right)
\end{split}
\end{equation}

\subsubsection{Balance Between operational and capital carbon Incentive} \label{sec:balance4}

\textit{Q.4: What level of incentive is required to align sustainable decisions with economic drivers?}
The effects of DC carbon incentive (e.g., carbon price) differ depending on whether they are applied at the operation site $w_{OP}$ or at the manufacturing site $w_{CA}$. Incentives in operations are reflected by weighting the carbon intensity $CI$ in the electricity price $EP$ to obtain a new electricity price $EP_I$. In contrast, CAPEX incentives are computed directly in the cost of the platforms:

\begin{equation}
\begin{split}
EP_I = EP + w_{OP} \times CI \\
C_{I,CA} = C_{F,CA} + w_{CA} \times C_{CFP,CA}
\end{split}
\end{equation}

Incentives applied to OPEX promote the adoption of new, more energy-efficient platforms. In contrast, CAPEX incentives discourage hardware replacement by increasing the cost of embodied emissions, thereby promoting the extended use of legacy systems. Therefore, $w_{OP}$ should be applied only in OPEX-dominated scenarios, and $w_{CA}$ only in CAPEX-dominated ones.

\subsubsection{Balance Between Acceleration and Energy Efficiency} \label{sec:balance5}

\textit{Q.5: How should performance acceleration and energy efficiency be balanced to achieve sustainability and viability?}
We model the joint impact of acceleration and energy efficiency on the total cost $C$ (in economic and emissions terms) of procuring an alternative platform \( d \) with unit cost of $C_{u,d}$ under the projected compute demand growth \( \eta \) as:
\begin{equation}
\label{eq:acc-eeff}
C = \left(\gamma \times \frac{E_{B}}{e} \times PUE + C_{u,d} \times \frac{N_{b}}{a}\right) \times \eta
\end{equation}

\noindent where $\gamma$ is the operational cost, quantified either economically ($EP$) or in emissions ($CI$). $N_{b}$ is the number of baseline platforms and $E_{B}$ the energy required to operate them.

This formulation decouples the effect of \textbf{acceleration} $a$ and \textbf{energy efficiency} $e$. By focusing on DC throughput rather than individual application latency, we enable flexible resource-time allocation across workloads. This approach directly ties the degree of acceleration to the number of platforms required to compute a given workload. Energy efficiency lowers energy consumption, thereby reducing OPEX. These two dimensions can present trade-offs: designing for higher acceleration may come at the expense of reduced energy efficiency, and vice versa.

\subsubsection{Balance Between Flexibility and Specialization} \label{sec:balance6}

\textit{Q.6: When should specialization be favored over flexibility from a sustainability and economic point of view?}
While specialized hardware can offer significant energy and performance gains for specific applications, its overall impact depends on its aggregate contribution to DC scheduling. Platform selection should therefore consider workload composition and execution dependencies. For platforms targeting multiple applications, we account for workload composition by defining the equivalent acceleration \( a_{eq} \) and energy efficiency \( e_{eq} \) across a set of accelerated applications \( K \) as:

\begin{equation} \label{eq:accel_eff}
a_{eq} = \frac{RTU_{K}}{\sum_{k \in K} \frac{RTU_k}{a_k}}, \qquad
e_{eq} = \frac{\sum_{k \in K} RTU_k \times P_k}{\sum_{k \in K} \frac{RTU_k \times P_k}{e_k}}
\end{equation}

\noindent
where \( RTU_k \) is the utilization of the application resource time \( k \), \( a_k \) and \( e_k \) are the acceleration and energy efficiency when running $k$, and \( P_k \) is its power use.

\subsection{Carbon and Price Efficiency Metrics}

To support decision-making around the six core questions posed in the model, we introduce
\textit{Carbon Efficiency} ($CE$, measured in FLOP/tCO$_2$-eq) and 
\textit{Price Efficiency} ($PE$, measured in FLOP/\$).
These metrics, defined in Table~\ref{tab:metrics_table}, depend on both platform-specific parameters (performance $Perf$, $Power$, capital cost per device $C_{F,u,d}$, and embodied carbon footprint per device $C_{CFP,u,d}$) and contextual variables ($CI$, $EP$, and device lifetime $T$).

\begin{table}[tp]
\caption{Carbon and economic efficiency metrics.}%
\label{tab:metrics_table}
\resizebox{0.47\textwidth}{!}{%
\begin{tabular}{@{}lcc@{}}
\toprule
\textbf{Metrics}          & \textbf{Carbon Efficiency ($CE$)} & \textbf{Price Efficiency ($PE$)} \\
                 & \textbf{[FLOP/tCO$_2$-eq]}        & \textbf{[FLOP/\$]} \\
\midrule
\textbf{Operational ($OP$)} & $CE_{OP} = \dfrac{Perf}{Power \times CI \times PUE}$ & $PE_{OP} = \dfrac{Perf}{Power \times EP \times PUE}$ \\
\midrule
\textbf{Capital ($CA$)} & $CE_{CA} = \dfrac{Perf \times T}{C_{CFP,u,d}}$ & $PE_{CA} = \dfrac{Perf \times T}{C_{F,u,d}}$ \\
\midrule
\textbf{Total}            & $CE = 1/\big(\dfrac{1}{CE_{OP}} + \dfrac{1}{CE_{CA}}\big)$ & $PE = 1/\big(\dfrac{1}{PE_{OP}} + \dfrac{1}{PE_{CA}}\big)$ \\
\bottomrule
\end{tabular}%
}

\end{table}

Using these metrics and Eq.~\ref{eq:CFP_growth}, we define the \textit{maximum sustainable growth} $\eta_S$ of an upgrade from a baseline device $d_B$ to an alternative device $d_A$ as the largest allowable increase in compute capacity under an iso-carbon policy:
\begin{equation}
\label{eq:max_growth}
\eta_{S} \leq \frac{CE(d_A)}{CE_{OP}(d_B)} \
\end{equation}

We can now define the \textit{carbon-economy gap} as the misalignment between carbon reduction and profit gains between two options. This gap informs policymakers on the \textit{incentives} required to stimulate the adoption of sustainable strategies through market instruments, taxes, subsidies, corporate commitments, or broader socioeconomic cost models.
Formally, the incentive $w$ is given by:

\begin{equation}
\label{eq:incentive}
w = \frac{\frac{1}{PE(d_B)} - \frac{1}{PE(d_A)}}{\frac{1}{CE(d_A)} - \frac{1}{CE(d_B)}}
\end{equation}

\begin{table}[th]
\caption{Simplified incentive expressions for Eq.~\ref{eq:incentive}.}%
\label{tab:incentive}
\resizebox{0.4\textwidth}{!}{%
\begin{tabular}{@{}lcc@{}}
\toprule
\textbf{Incentives}          & \textbf{Expression} \\
\midrule
\textbf{Upgrades} & $CE(B) = CE_{OP}(B)$; $PE(B) = PE_{OP}(B)$ \\
\midrule
\textbf{OPEX-dominated}         & $CE(A) = CE_{OP}(A) $ \\
\midrule
\textbf{CAPEX-dominated}            & $CE(A) = CE_{CA}(A) = 0 $ \\
\bottomrule
\end{tabular}%
}

\end{table}

Table~\ref{tab:incentive} shows the simplified expressions of Eq.~\ref{eq:incentive} for upgrades, OPEX-dominated, and CAPEX-dominated scenarios.

\section{Analysis: CEO-DC for HPC AI Case Study} \label{sec:results}

This case study demonstrates the real-world applicability of the \ecodc framework in specialized AI DCs.
Models like GPT, LLaMA, and Meditron require vast amounts of dedicated servers running for weeks or months to complete a single training run~\cite{faiz_llmcarbon_2023, chen2023meditron}.
For this reason, we focus on monolithic workloads and use the CEO-DC framework in AI and HPC.
We begin by quantifying emissions and economic factors as inputs to the framework.
The framework is then used to:  
(i) evaluate the carbon-economy trade-offs from our model on top-tier AI benchmarks and across leading global DC regions, and  
(ii) showcase the application of carbon and price metrics to guide DC planning in the current AI platform landscape.
Throughout this analysis, we highlight the findings on actionable insights for DC stakeholders.

\subsection{Methodology for Quantifying Emissions and Economic Factors}
\label{sec:experimental_setup}

We use the MLPerf AI Training and HPC Inference Benchmarks~\cite{mattson2020mlperf, MLPerf_HPC_2021} to analyze performance and energy consumption trends across hardware generations. These benchmarks provide standardized, cross-vendor metrics that are well-suited for evaluating carbon-aware upgrade strategies in DCs focused on AI. Our analysis includes MLPerf versions v0.7 to v3.0 (Closed TTT) for HPC inference, and v0.7 to v5.0 (Closed and Closed-Power) for AI training. We consider server-grade platforms from 2018 to 2024, including NVIDIA (V100, A100, L40S, H100, B200, H200), Google (TPU v3 to v6), Huawei (Ascend910a), Cerebras (GC200), Habana/Intel (Gaudi2), and Graphcore (Bow). Google TPUs are only considered baselines as they are not commercially available for procurement. In the absence of measured energy data, we use thermal design power (TDP) as a proxy for energy consumption, following prior work~\cite{patterson2021carbon, faiz_llmcarbon_2023}. We select the submissions with the lowest per node latency and energy for each benchmark and platform release year. For reproducibility, the selected benchmarks are listed in the annex.

\begin{figure}[!tp]
    \centering
    \includegraphics[width=1\linewidth]{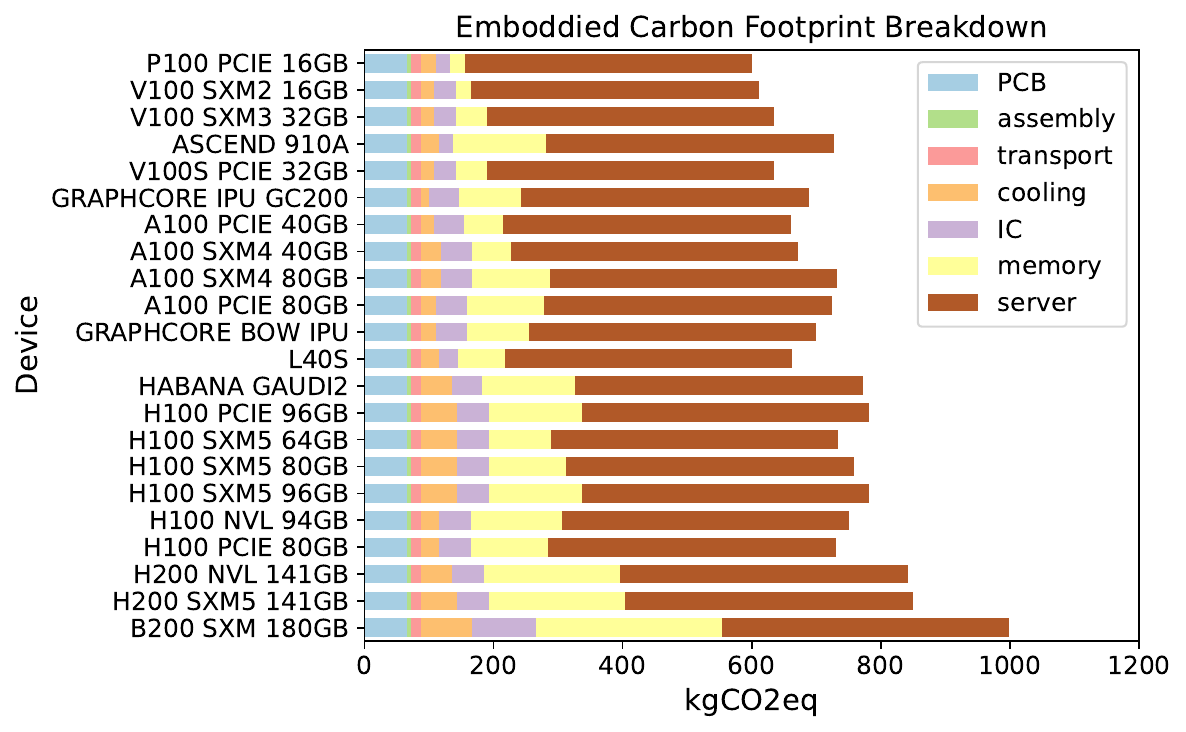}
    \caption{Embodied emissions for HPC platforms.}
    \label{fig:stacked_cfp}
\end{figure}

Fig.~\ref{fig:stacked_cfp} shows the breakdown of the embodied emissions of each platform used in our study, calculated using Eq.~\ref{eq:embodied} with the following data sources:
IC manufacturing and packaging ($C_{IC,d}$) from the IMEC NetZero tool~\cite{IMECNetZero_2025} (details in Table~\ref{tab:platforms} from the Annex);
\SI{66.10}{\text{kgCO$_2$-eq}} for PCB manufacturing ($C_{mb,d}$) and \SI{6.68}{\text{kgCO$_2$-eq}} for assembly ($C_{asm,d}$), both from Boavizta~\cite{Boavizta_2025};
transportation to the DC ($C_{tp,d}$) estimated using CarbonCare.org (ISO 14083:2023) assuming \SI{2}{kg} per platform (values per location in Table~\ref{tab:countries} from the Annex);
and cooling system impact ($C_{cool,d}$) based on Gupta et al.~\cite{li_LLM_LCA_2024}, which uses a \SI{0.0788}{\text{kgCO$_2$-eq}/W} multiplied by the TDP value.
Memory production ($C_{mem,d}$) is particularly difficult to assess, as no open LCA database exists for DRAM technologies. We therefore adopt a conservative estimate of \SI{1.5}{\text{kgCO$_2$-eq/GB}} for all DRAM types from Boavizta~\cite{Boavizta_2025} based on Samsung Technologies (the most pessimistic figures in the literature).

Values dependent on server configuration are based on a typical setup: \SI{1780}{\text{kgCO$_2$-eq}} for a Lenovo SR675 V3 Base Module amortized across four PCIe/SXM5 slots~\cite{lenovo2025carbonfootprint}. Server replacement constitutes a substantial portion of the embodied emissions, often driven by the need to support newer components. However, not all subsystems---such as SSDs, hard drives, or even CPUs---always require replacement.

\begin{tcolorbox}[title=Actionable Insights]
\textbf{Platform Designers:} A modular design that maintains compatibility with new accelerators can avoid unnecessary replacements, potentially saving between \SI{50}{\percent} and \SI{75}{\percent} of embodied emissions, depending on the platform. This might not always be possible, as in the case of the PCIe, which is integrated in the IC of the CPU and has been improving every few years.
\end{tcolorbox}

The economic cost is tracked using market release prices, adding \SI{3500}{\$} to a Lenovo SR675 V3 Base Module for every four devices. Based on historical data for NVIDIA GPUs, which have been leading market performance, we have empirically estimated an annual exponential increase in GPU prices of approximately \SI{24}{\percent}. For reference, we have used the following platform specifications: V100 SXM2 16GB (2017, \$9500), A100 SXM4 80GB (2020, \$18500), H100 NVL 94GB (2023, \$33000), and B200 SXM 180GB (2024, \$44250). Similarly, we empirically estimate an exponential increase in embodied emissions per platform of approximately \SI{4}{\percent/year}.

The gross income value \( G_B \) is estimated based on the equivalent cost of running the applications on an Nvidia H100, using a cloud price of \SI{2.65}{\$/h} as a reference. We extract the CI per country from Our World in Data~\cite{ember_energyinstitute_carbon_intensity_2025} and the projected electricity price ($\beta$) from Electricity Maps~\cite{electricitymaps2025} (values per country summarized in Table~\ref{tab:countries} in the Annex). Finally, we assume full node utilization and a PUE of \SI{1.25}{}.

We have empirically estimated DC capacity growth rate by combining evidence from the IEA~\cite{IEAelec2024} and the Top500 list~\cite{top500_perfdevel}. Both sources indicate sustained exponential growth in DC compute capacity. From the Top500, we use the FLOP/s reported per year, and obtain yearly compute capacity growth rates of \SI{40}{\percent} for the \#500 system and \SI{49}{\percent} for the top-ranked system. In parallel, the IEA reports the energy consumption of AI workloads in DCs. We translate these energy projections into capacity growth trends using $ \eta = e_{\text{avg}} \times E_{\text{trend}} $ where $e_{\text{avg}}$ is the average energy efficiency improvement and $E_{\text{trend}}$ the projected energy demand growth. This yields annual capacity growth rates of \SI{55}{\percent}, \SI{60}{\percent}, and \SI{66}{\percent} under the IEA's Headwinds, Base, and Lift-off scenarios, respectively.

\subsection{Guiding DC Decisions Toward Long-Term Sustainability}

Here, we first analyze the carbon-economy model and its questions through current platform design trends, increase in DC capacity, and compatibility of platform upgrades with sustainable growth. We then apply the same analysis to the context of different countries to give a global overview.

\subsubsection{Following the Trends}

\begin{figure}[htbp]
    \centering
    \includegraphics[width=\linewidth]{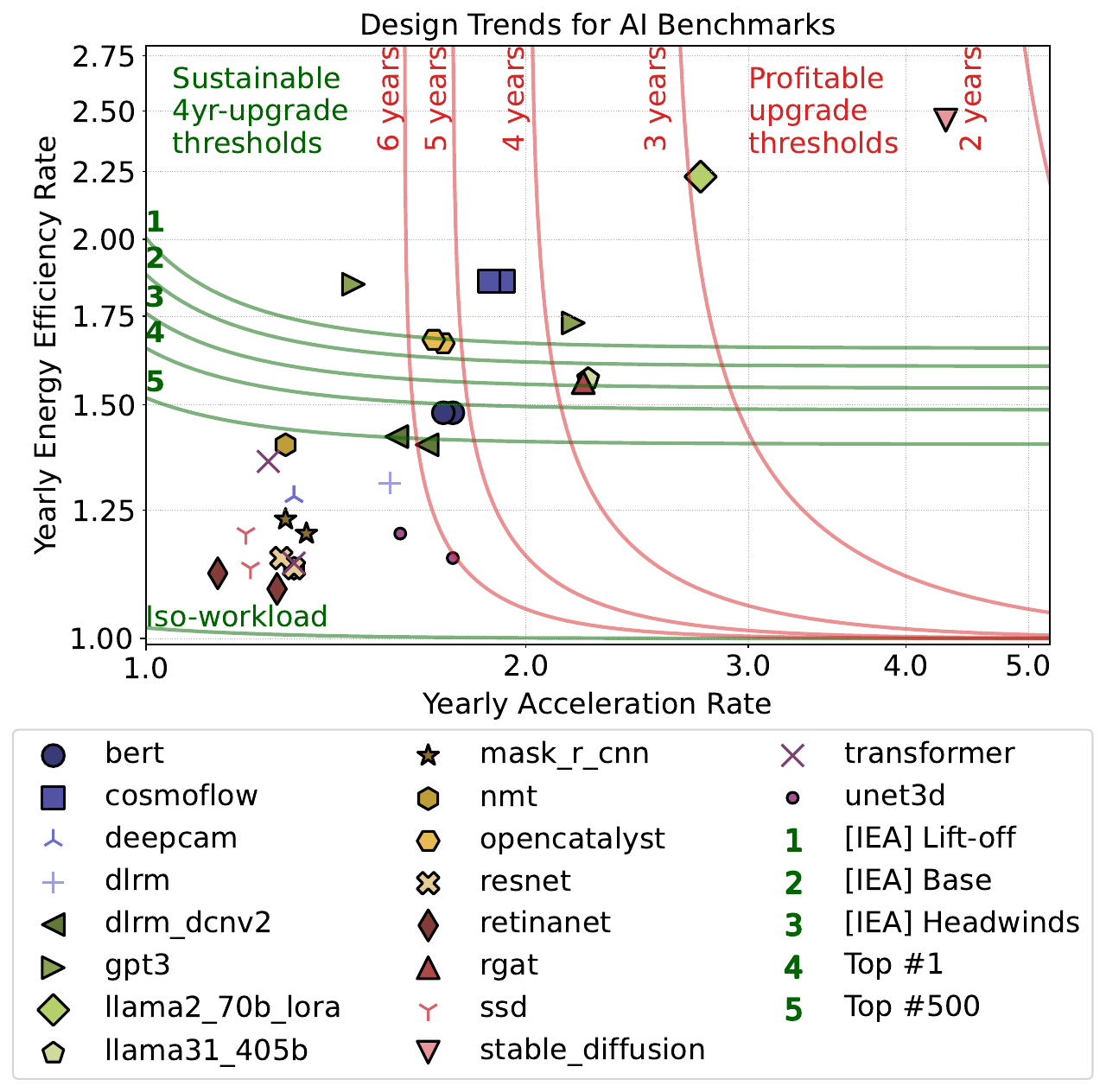}
    \caption{AI platform trends from an Nvidia H100.}
    \label{fig:dse_platform_trends}
\end{figure}

We employ CEO-DC to analyze future platform development according to current trends. Fig.~\ref{fig:dse_platform_trends} shows the performance and energy efficiency improvement trends for each benchmark. Each benchmark presents one point corresponding to the trends obtained with the fastest executions per year, and another one with the most energy efficient. The sustainability and viability thresholds (green and red lines, respectively) are obtained from Eq.~\ref{eq:acc-eeff}, using an NVIDIA H100 baseline. We use the world average \SI{400}{\text{gCO$_2$-eq/kWh}} CI and \SI{0.15}{\$/kWh} electricity price. The rate of improvement varies significantly by benchmark, and it is influenced by how well hardware resources are utilized. On average, energy efficiency improvements lag behind performance improvements, with a yearly improvement rate \SI{18}{\percent} slower---limiting the ability to reduce operational emissions. In addition, only a few benchmarks reveal notable trade-offs between top performance and top energy efficiency deployments---such as \texttt{gpt3} and \texttt{retinanet}.

We apply the carbon-economy model to address the trade-offs of the DC decision space. To this end, we plot the sustainable and viable upgrade thresholds from Eq.~\ref{eq:acc-eeff}---green and red lines, respectively.
All benchmarks fall above the four-year upgrade cycle sustainability threshold for iso-workload, meaning that OPEX savings outweigh CAPEX emissions (Q.1).
However, there is a carbon-economy misalignment in \SI{72}{\percent} of benchmarks by not being naturally incentivized for less than four-year cycles, and \SI{50}{\percent} do not reach the six-year cycle threshold Q.2).
Moreover, only \SI{28}{\percent} of benchmarks can reduce emissions while sustaining a capacity increase of \SI{6.63}{\times/year} derived from the IEA Base Case (Q.3).

These findings also transfer to the design space of platforms.
Sustainable computing in HPC and AI DCs is mainly driven by Operational Carbon Efficiency ($CE_{OP}$), but current improvements in platform energy efficiency alone are insufficient to offset the growing demand in most cases. Consequently, decarbonization of electricity generation is essential to achieve sustainable growth. However, the economic viability of sustainable upgrades is largely determined by Capital Price Efficiency ($PE_{CA}$). Despite platform acceleration rates, current platform prices discourage the timely replacement of inefficient hardware.

\begin{tcolorbox}[title=Actionable Insights]
\textbf{Platform Designers:}
Designs must optimize both $CE_{OP}$ and $PE_{CA}$ to enable sustainable computing with affordable upgrades (Q.5). 
Furthermore, application-specific hardware development should target mainstream workloads that do not show significant improvements across GPU generations (Q.6).

\medskip
\textbf{DC Managers:} Upgrade current GPUs to next-generation dedicated platforms that demonstrate substantial improvements for DC benchmarks (e.g., \texttt{llama}, \texttt{gpt3}), while leaving legacy platforms to absorb the demand of applications that do not exhibit such improvements (e.g., \texttt{resnet}) or to domains with lower utilization rates (e.g., cloud).
\end{tcolorbox}

Interestingly, current trends point to divergent trajectories for OPEX and CAPEX emissions. On the operational side, OPEX emissions are projected to rise, as annual improvements in energy efficiency (\SI{43}{\percent}) fall short of the growth in compute capacity (\SI{60}{\percent}). In contrast, CAPEX emissions are expected to decline, given that the annual speedup rate (\SI{74}{\percent}) exceeds capacity growth.

\begin{table}[hbp]
\caption{Compute demand growth in 4 years according to different scenarios and scopes. A design incentive (in \$/tCO$_2$-eq) is required when the projected compute demand growth exceeds the growth enabled by design trends.}
\label{tab:compute-growth}
\resizebox{0.47\textwidth}{!}{%

\begin{tabular}{@{}cccc|cc@{}}
\toprule
Scenario & Demand & Acc. & Energy Eff. & \multicolumn{2}{c}{Design Incentive} \\
         & Growth & Gain & Gain        & Total & OPEX-only \\
\midrule
Top \#500                & 3.86 & 9.69 & 3.96 & 0 & 0 \\
\textbf{Design Trends} & \textbf{4.1} & \textbf{9.17} & \textbf{4.18} & \textbf{0} & \textbf{0} \\
Top \#1 (in Top \#500)                  & 4.90 & 7.51 & 5.11 & 3199 & 3072 \\
{[}IEA{]} Headwinds Case & 5.70 & 6.35 & 6.03 & 3881 & 3699 \\
{[}IEA{]} Base Case      & 6.63 & 5.34 & 7.18 & 4737 & 4473 \\
{[}IEA{]} Lift-off Case  & 7.52 & 4.58 & 8.37 & 5642 & 5276 \\
\bottomrule
\end{tabular}
}

\end{table}

We observe that AI HPC platforms are naturally incentivized to be replaced every six years on average, considering the average improvement rates in performance and energy across benchmarks. However, due to the large share of operational emissions in HPC and AI, shortening replacement cycles can reduce total emissions by early substituting inefficient platforms. For this purpose, our policy space model suggests applying OPEX-only incentives to promote earlier replacements (Q.4). To justify sustainable upgrades on a five-year cycle, an additional \SI{87}{\text{USD/tCO$_2$-eq}} incentive in operations would be required on average, and \SI{512}{\text{USD/tCO$_2$-eq}} for a four-year cycle. For context, the current global average carbon price is \SI{32}{\text{USD/tCO$_2$-eq}}, with the highest price applied being \SI{167}{\text{USD/tCO$_2$-eq}} in Uruguay~\cite{WorldBankCarbonPricing2025}.
In conclusion, promoting shorter upgrade cycles implies moving further away from profitability-driven replacement trajectories and requires increasingly large financial support.

Economic incentives also shape the development of platforms. We analyze the incentives required to promote designs that could sustain the current computing growth trend. Table~\ref{tab:compute-growth} presents the minimum platform-level improvements required---in energy efficiency and performance---to meet projected computing growth scenarios while maintaining carbon neutrality. Considering the benchmark-dependent nature of the performance-energy trade-off, we illustrate feasible design options with the same energy-delay product (EDP). Our results indicate that prioritizing energy efficiency at the cost of latency to achieve sustainable scaling can lead to prohibitively high carbon prices. For example, justifying a platform capable of scaling to the IEA Base Case demand growth scenario would require an additional incentive of \SI{4450}{\text{USD/tCO$_2$-eq}}.

\begin{tcolorbox}[title=Actionable Insights]
\textbf{Policymakers:} Carbon pricing on the grid can incentivize both shorter upgrade cycles and the development of carbon-efficient platform designs, bridging the carbon-economy gap in sustainable computing. 

\medskip
\textbf{Platform designers:} Current design trends require prohibitively high carbon incentives to accelerate the replacement of inefficient hardware. A fundamental shift improving both performance and energy efficiency is needed to achieve sustainable growth supported by frequent, viable upgrade cycles.
\end{tcolorbox}

\subsubsection{Regional Insights and Comparative Analysis}

\begin{figure*}[tp!]
    \centering
    \includegraphics[width=0.95\linewidth]{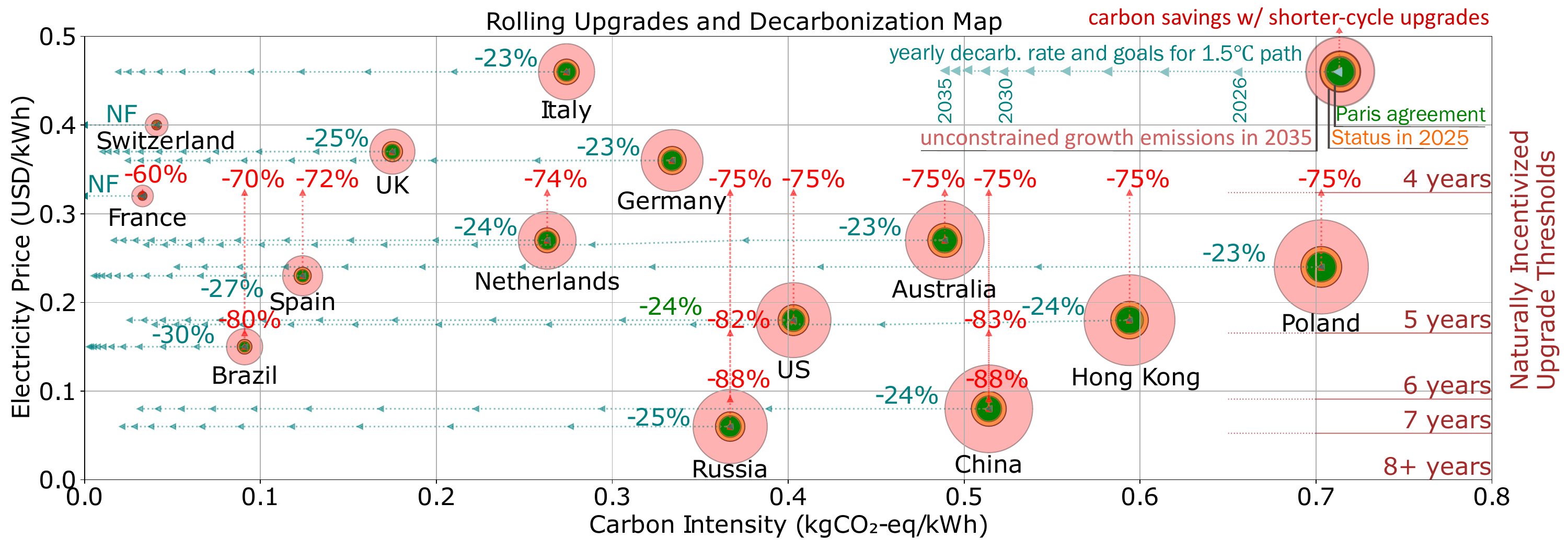}
    \caption{Upgrade policies and decarbonization needs by country for current platform improvement trends and compute demand growth. 
    }
    \label{fig:multiplatform_optimization_caps_loc}
\end{figure*}

In this section, we apply the proposed framework at a global scale, evaluating regional dependencies for platform procurement upgrade cycles and grid decarbonization requirements given the ongoing demand. In this context, the 2024 UN Emissions Gap Report~\cite{unep_emissions_gap_report_2024} calls for a \SI{57}{\percent} global emissions reduction by 2035 to realign with the \SI{1.5}{\degreeCelsius} Paris Agreement pathway. Fig.~\ref{fig:multiplatform_optimization_caps_loc} summarizes how two orthogonal strategies---grid decarbonization (blue) and timing upgrade cycles (red)---depend on regional variations in carbon intensity ($x$-axis) and electricity prices (left $y$-axis). Three concentric circles represent countries with the highest number of DCs in 2025~\cite{Statista2024DataCentersByCountry}. The orange circle is proportional to the operational emissions, assuming the same workload in all regions. The green circle defines the Paris Agreement emission reduction target for 2035, and the red circle represents the projected emissions for 2035 with current unregulated growth. Each blue arrow shows the reduction of CI that must be achieved in subsequent years from 2024 if the constant yearly decarbonization rate target (blue number) is followed. The upgrade cycles that each region is economically incentivized to follow depend solely on the electricity price and are marked on the right $y$-axis (4, 5, 6, 7, and 8+ years). The length of the vertical red arrows corresponds to the increase in the electricity price (or an alternative carbon incentive) required to promote shorter upgrade cycles. Finally, the red number shows potential carbon savings from timely upgrades. For instance, replacing in Poland 4-year-old platforms for B200 could cut their associated emissions by \SI{75}{\percent}.

This analysis exposes the excessively high decarbonization rate needed (from \SI{-23}{\percent/year} to \SI{-30}{\percent/year} in all countries) to meet the UN emissions goal without compromising the growth of the compute capacity. The required decarbonization rate is even steeper in countries with low CI, highlighting the unequal challenges faced by those further along in their decarbonization journey. In these regions, CAPEX-dominated emissions reveal diminishing returns of additional decarbonization efforts. France and Switzerland exemplify this case, where infrastructure replacement or expansion would generate CAPEX-related emissions surpassing their targets, resulting in non-feasibility (NF). The decarbonization rate is derived by simulating rolling upgrades from 2024 to 2035, annually substituting platforms beyond the upgrade threshold, and replacing them with the latest generation to accommodate demand growth. We start with a server fleet of B200 nodes and assume global trends from Sect.~\ref{sec:experimental_setup}.

Across all regions, economic incentives favor DC upgrade cycles of four years or longer. However, all regions could reduce total emissions by shortening their upgrade cycles---based on performance improvement trends for AI platforms---due to their OPEX-dominated emissions. Moreover, some low-CI regions with less potential for carbon savings (e.g., Switzerland and the UK) are incentivized to keep short-cycle upgrades due to their high electricity price. Three-year upgrade cycles---or shorter---are prohibitively expensive worldwide, with electricity price thresholds exceeding \SI{0.71}{\$/kWh}. In regions with high-CI, replacing 4-year-old platforms today yields to \SI{75}{\percent} reduction of their total associated emissions, \SI{80}{\percent} for 5-year-old platforms, and \SI{88}{\percent} for 6-year-old ones. 

This analysis assumes energy efficiency improvement trends for specialized AI HPC platforms, which are expected to be higher than the improvement trends for general-purpose platforms. Using the CEO-DC model, we find that the optimal upgrade policy changes depending on the energy efficiency rate assumed. For instance, France should only upgrade 4-year platforms in 2024 if the energy efficiency takes less than \SI{63}{months} to double; \SI{75}{months} for Switzerland. This implies that ResNet needs 5-year upgrade cycles in these two countries. A table with the minimum energy efficiency rate per region is shown in the Annex for reference. Based on our model, electricity price thresholds scale with platform price and energy efficiency trends, and inversely with performance trends, yielding a \SI{1.9}{\percent} annual increase for our use case. This rate, together with inflation, could play a decisive role in shaping future upgrade policies.

\begin{tcolorbox}[title=Actionable Insights]
\textbf{DC Managers:} Shortening upgrade cycles can reduce total carbon emissions, but must be paired with aggressive grid decarbonization or controlled capacity growth to remain aligned with sustainability targets.

\medskip
\textbf{Policymakers:} To ensure equitable and effective global action, sustainability goals must be distributed fairly between regions. This approach allows all countries to contribute meaningfully while maintaining room for technological and economic development.
\end{tcolorbox}

\subsection{DC Planning using CEO-DC metrics}

This section presents results from applying the CEO-DC framework to specific cases to analyze the carbon-economy gap across the commercial AI platform landscape.

\subsubsection{Planning in Current AI Platforms Landscape}

\begin{figure}[!tp]
    \centering
    \includegraphics[width=\linewidth]{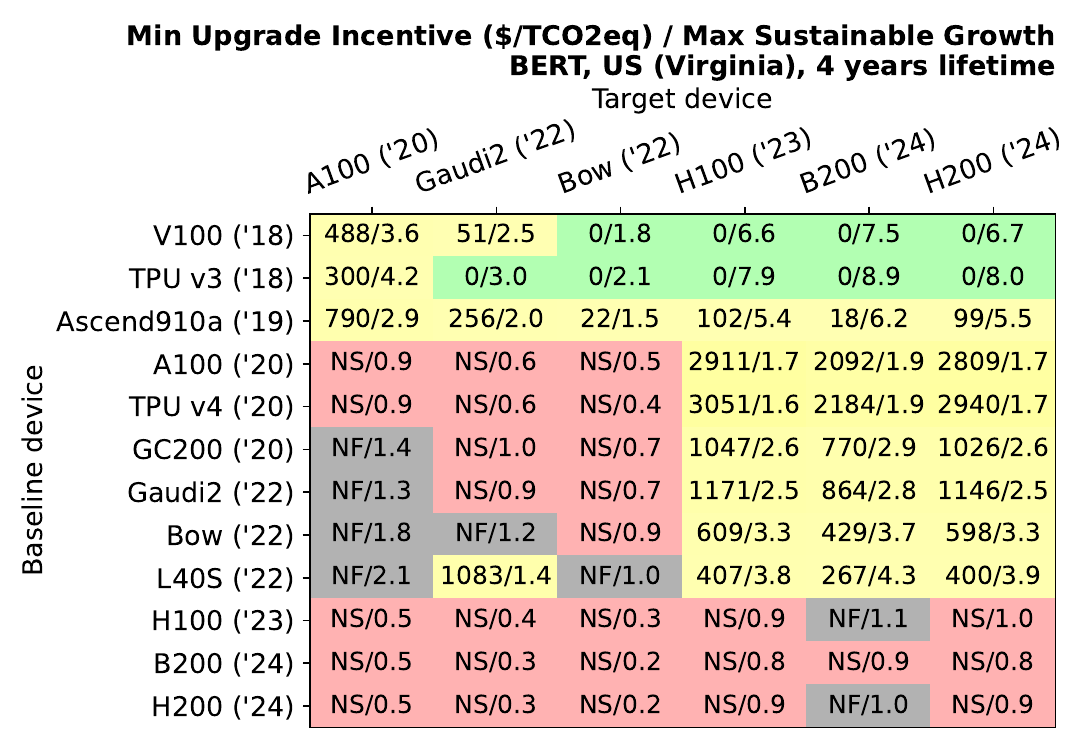}
    \caption{Minimum upgrade incentive and maximum sustainable growth in a 4-year cycle for platforms profiled with BERT~\cite{mattson2020mlperf}. We show naturally incentivized viable upgrades (green), sustainable upgrades requiring incentives (yellow), non-sustainable upgrades (red, NS), and upgrades not economically feasible to incentivize (grey, NF).}
    \label{fig:incentive_growth_table}
\end{figure}

We apply CEO-DC metrics to demonstrate their role in DC planning. To this end, Fig.~\ref{fig:incentive_growth_table} illustrates all possible profitable platform upgrades from a given baseline (left axis) to any possible target device (upper axis). Platform comparisons are based on BERT for its extensive standardized profiling coverage in the MLPerf benchmark suite~\cite{mattson2020mlperf}. We use the contextual data from Virginia (US) due to its high concentration of DCs. In this figure, we evaluate the \textit{minimum incentive} required to promote each upgrade, and the \textit{maximum sustainable growth} achievable within iso-carbon constraints. Upgrades that are sustainable but not economically viable fall into either incentivized (yellow) or non-feasible (grey) categories. A solution is no longer feasible to incentivize when the required subsidy increases total costs beyond the gross income, resulting in a scenario where non-operation is more economically favorable.

Our analysis reveals key trends in platform replacement. Replacing platforms released in 2018 with 2022 platforms is naturally incentivized and sustainable. In contrast, replacing 2020 with 2024 platforms requires substantial carbon incentives. In terms of sustainable compute growth, replacing a V100 with an A100 over a four-year cycle allows for up to a \SI{3.6}{\times} growth under iso-carbon constraints, while upgrading from an A100 to a B200 only supports a \SI{1.9}{\times} increase. 

The \textit{sustainable growth} enabled by the upgrade depends on the target workload. The same analysis for LLaMA2\_70b\_lora reveals that upgrading the A100 is both sustainable and naturally incentivized when transitioning to the H100 (2023) or B200 (2024), enabling sustainable compute growth of \SI{11.8}{\times} and \SI{20.5}{\times}, respectively. In contrast, upgrading from a V100 to an A100 in ResNet is sustainable but not feasible to incentivize. Similarly, upgrading from an A100 to an H100 lacks sufficient performance improvement. Even a V100-to-H100 transition would require an unrealistically high carbon price of \SI{2401}{USD}/TCO\(_2\)-eq to be economically justified.

\begin{tcolorbox}[title=Actionable Insights]
\textbf{DC Managers:} Carbon efficiency informs about the sustainable scaling potential of a platform, while its combination with Price Efficiency reveals whether the upgrade is also economically justified. Considering both metrics is key to fully evaluating the economic and environmental implications of DC procurement decisions.
\end{tcolorbox}

\subsubsection{Exploring DC Procurement Strategies and Incentives}

\begin{figure}[!t]
    \centering

    \begin{subfigure}[]{0.4\textwidth}
        \includegraphics[width=\linewidth]{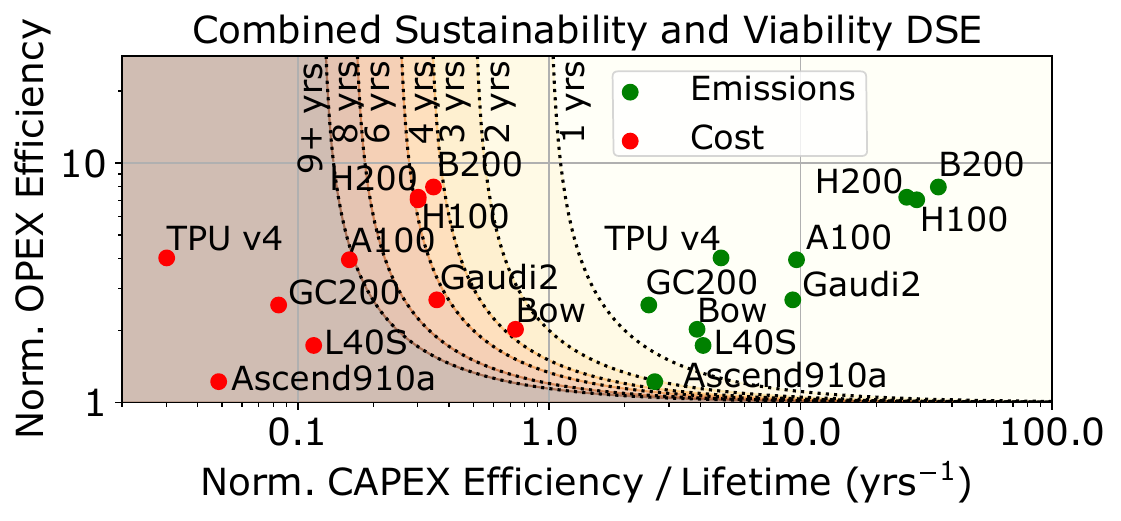}
        \caption{OPEX vs. CAPEX efficiency metrics.}
        \label{fig:dse_opex_capex}
    \end{subfigure}
    \hspace{0.02\textwidth}
    \begin{subfigure}[]{0.4\textwidth}
        \includegraphics[width=\linewidth]{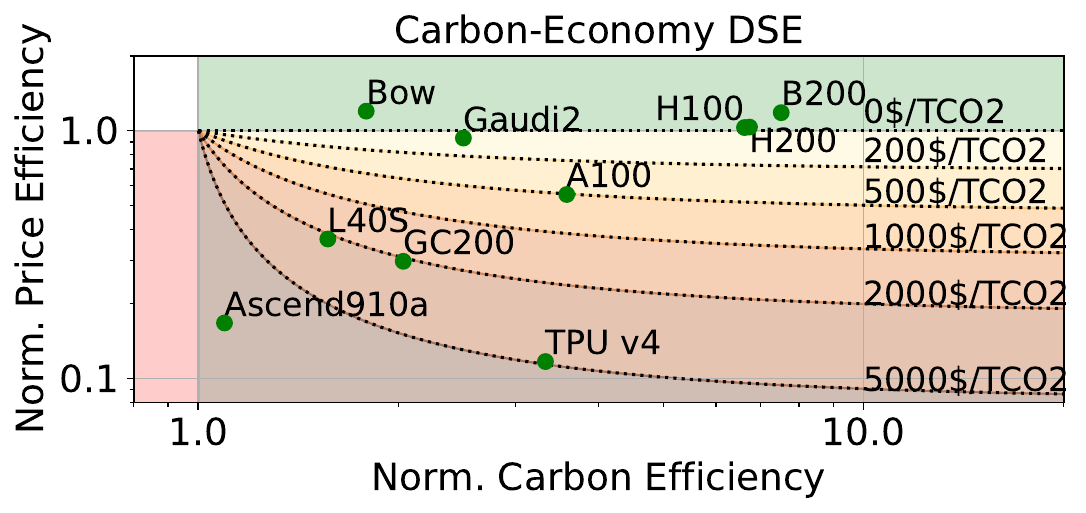}
        \caption{Four-year lifetime Price vs. Carbon Efficiency.}
        \label{fig:dse_inceintive}
    \end{subfigure}%

    \caption{Carbon and Price Efficiency for HPC AI platforms from 2016 to 2024 for MLPerf~\cite{mattson2020mlperf} BERT benchmark. All metrics are normalized to the operational efficiency of the baseline (i.e., $CE_{OP}(d_B)$ for emissions and $PE_{OP}(d_B)$ for costs).}
    \label{fig:dse_metrics}
\end{figure}

Carbon and price efficiency metrics ($CE$, $PE$) are useful for timing platform upgrades.
Fig.~\ref{fig:dse_metrics} presents OPEX, CAPEX, and total $CE$ and $PE$ normalized to a V100. 
Figure~\ref{fig:dse_opex_capex} separates OPEX and CAPEX efficiencies for costs and emissions---normalized to the OPEX from V100. We analyze different amortization thresholds by representing the CAPEX efficiency per lifetime ($x$-axis).
The figure reveals that while the newer platforms deliver carbon reduction benefits, $PE_{CA}$ undermines the return on investment of upgrading. 
For example, upgrading to A100 is only economically viable when replacing the V100 if amortized over more than eight years. This drops to four years only with the introduction of the Bow.  
The analysis in Fig.~\ref{fig:dse_inceintive} shows that the $PE$ for many platforms is insufficient (<1, brown gradient area) to justify the upgrade when amortized over four years. For instance, the A100 would require an incentive of \SI{500}{\text{USD/tCO$_2$-eq}}.

These metrics shape the optimal decision in DC planning. The Bow is the most profitable platform to acquire, with a $PE$ slightly above the B200 (Fig.~\ref{fig:dse_inceintive}, green area). However, the B200 has \SI{4}{\times} more carbon efficiency than the Bow. This implies that the B200 enables four times more capacity growth with the same carbon budget or cap. Specifically, substituting the V100 for the B200 enables \SI{6}{\times} capacity growth at iso-carbon.
This comparison highlights the contrast between cost- and carbon-efficient platform selection strategies. Although Bow provides immediate economic gains for 4-year amortization, the B200 is more suitable for sustainable scaling due to its carbon efficiency.
For larger amortization periods, the B200 would remain the most price and carbon-efficient---due to its $PE_{CA}$ increase.
Moreover, we could align economic interests with carbon-efficient decisions regardless of the carbon budget by applying an incentive. \SI{13}{\text{USD}} per tCO\(_2\)-eq of total emissions would be required to economically favor the B200 over the Bow for a 4-year amortization.

The profitability of the computation determines the price point at which a DC manager is willing to expand.
Under a carbon-capped scenario, indefinite growth would require a penalty of $\approx$\SI{7000}{\text{USD/tCO\(_2\)-eq}} to enforce any desired emissions limit. Other platforms prior to Bow (2022) lack the price efficiency required to justify their update, as illustrated by the incentives of Fig.~\ref{fig:dse_inceintive}. Similarly, Fig.~\ref{fig:dse_opex_capex} shows that the $PE_{CA}$ mainly constrains the viability of platforms---insufficient acceleration and excessively high prices of platforms.

\begin{tcolorbox}[title=Actionable Insights]
\textbf{Policymakers:} Incentives can be used to pivot the upgrade decision towards carbon-efficient platforms, offering continued emissions savings as demand grows. To be effective, emission caps should be reinforced with penalties above the incentives of unconstrained growth.

\medskip
\textbf{DC Managers:} Delay upgrades until improved carbon efficiency ensures sustainable growth. This may reduce short-term profitability but is critical for meeting carbon reduction goals.
\end{tcolorbox}

\section{Discussion: Limitations and Future Work} \label{sec:discussion}

\textbf{Domain Assumptions and Applicability.}
Our analysis focuses on dedicated HPC DCs, where fixed resource allocation and high utilization justify frequent upgrades through operational savings. This introduces a key limitation for direct use in cloud services, since sustainability and viability thresholds depend on knowing which applications will be scheduled. In addition, future work should incorporate QoS-driven profit models at the workload level for the cloud domain.
Furthermore, the results of our analysis do not apply directly to the mobile domain where CAPEX emissions dominate and the sustainability benefits of upgrades diminish~\cite{Gupta_2020}.

\textbf{Performance Trends and Hardware Lifetimes.}
Our assumptions are based on empirical exponential trends in compute performance and energy efficiency for server-grade AI platforms from 2018 to the present. Specifically, improvements in AI platforms have been possible due to the specialization of hardware architectures, memories, and communications. While these trends have held in recent years, a significant slowdown in hardware improvements would require extending platform lifetimes.

\textbf{Geopolitical Coordination and Global Equity.}
While cloud providers increasingly adopt carbon-aware workload shifting~\cite{MicrosoftSustainability2024, Radovanović_2021}, constraints in data sovereignty, regulatory requirements, and latency prevent this strategy from being universally applicable. Our analysis focuses on non-migratable workloads in the context of the local region. Standardizing global incentive frameworks could promote more equitable and effective progress toward emissions reduction targets while allowing regional adaptation and harnessing the potential to further improve sustainability by migrating workloads.

\textbf{Full Environmental Impact.} The conclusions of this paper are derived solely from the perspective of greenhouse gas (GHG) emissions, critical to the urgent challenge of climate change. However, prioritizing short-cycle upgrades based only on GHG reductions can exacerbate other environmental threats within the planetary boundaries of the Earth~\cite{richardson2023earth}, including increased water consumption, depletion of mineral resources, change of land use, and generation of hazardous waste.
A comprehensive assessment integrating these impact categories and their relative urgency across regions is needed to avoid shifting burdens from global climate impacts to equally critical local pressures. This work thus lays the groundwork for future research incorporating multi-criteria environmental evaluations into upgrade cycle decisions.

\section{Conclusion} \label{sec:conclusion}

We introduced \textsc{CEO-DC}, a holistic framework for carbon and economy optimization in DCs to address key gaps in the literature: the lack of progressive scalability considerations, carbon-aware platform design, and incentive mechanisms for sustainability. 
Our analysis for the HPC AI case study shows that platform upgrade cycles can be reduced to mitigate carbon emissions, but they need incentives to become viable. Worldwide, upgrade cycles shorter than 4 years are prohibitively expensive. In addition, a fundamental shift in platform design, carbon policies, and grid decarbonization efforts is needed to meet the computed demand growth projections while remaining within carbon budgets.
Ultimately, \textsc{CEO-DC} provides actionable insights for key stakeholders.
\textbf{DC managers} are encouraged to enforce timely upgrades, limit capacity growth, favor carbon efficiency in platform selection, and invest in the design of more efficient platforms. \textbf{Policymakers} can implement emission caps, provide incentives to influence platform choice, promote timely upgrades, and guide design trends toward sustainability. Finally, \textbf{platform designers} can specialize and optimize carbon and price efficiency, and make modular designs.

\section*{Acknowledgments}
This research was partially funded by the Swiss NSF, grant no.~200021E\_220194: ``Sustainable and Energy Aware Methods for SKA (SEAMS)'',
the UrbanTwin project (ETH Board Joint Initiatives for the Strategic Area Energy, Climate and Environmental Sustainability, and the Strategic Area Engagement and Dialogue with Society),
by the Swiss State Secretariat for Education, Research, and Innovation (SERI) through the SwissChips research project,
and supported by the EPFL EcoCloud Center.

\sloppy
\normalsize
\input{main.bbl}

\section*{ANNEX: Data for Reproducibility of Results}
\label{sec:annex_repro}

\begin{table*}[ht]
\caption{Embodied emissions for IC and memory oper device. The memory size, technology, and area are extracted from TechPowerUp GPU Database~\cite{techpowerupGPU}, and the embodied emissions are extracted from \cite{IMECNetZero_2025}. We assume squared areas, binomial yield function with defect density of \SI{0.1}{\text{defects/cm$^2$}}, IPCC Tier 2 with Combustion, with the CI from Taiwan. Missing data (e.g., GC200, Bow) was filled using official vendor documentation.}
\label{tab:platforms}
\resizebox{\textwidth}{!}{%
\begin{tabular}{lllllll}
\hline
\textbf{Model}               & \textbf{Memory Size} & \textbf{Memory} & \textbf{Technology} & \textbf{Area}  & \textbf{IC Embodied Carbon} & \textbf{Mem Footprint} \\
               & (GB)        &   Type & Node       & (mm2) & (kgCO2-eq) & (kgCO2-eq) \\
\hline
P100 PCIe           & 16     & HBM2   & N14   & 610.0      & 20.4  & 24     \\
V100 SXM2           & 16     & HBM2   & N14   & 815.0      & 33.9  & 24     \\
V100 SXM3           & 32     & HBM2   & N14   & 815.0      & 33.9  & 48     \\
V100s PCIe          & 32     & HBM2   & N14   & 815.0      & 33.9  & 48     \\
A100 PCIe           & 40     & HBM2e  & N7    & 826.0      & 47.3  & 60     \\
A100 PCIe           & 80     & HBM2e  & N7    & 826.0      & 47.3  & 120    \\
A100 SXM4           & 40     & HBM2e  & N7    & 826.0      & 47.3  & 60     \\
A100 SXM4           & 80     & HBM2e  & N7    & 826.0      & 47.3  & 120    \\
H100 PCIe           & 96     & HBM3   & N5    & 814.0      & 49.5  & 144    \\
H100 NVL            & 94     & HBM3   & N5    & 814.0      & 49.5  & 141    \\
H100 PCIe           & 80     & HBM3   & N5    & 814.0      & 49.5  & 120    \\
H100 SXM5           & 64     & HBM3   & N5    & 814.0      & 49.5  & 96     \\
H100 SXM5           & 80     & HBM3   & N5    & 814.0      & 49.5  & 120    \\
H100 SXM5           & 96     & HBM3   & N5    & 814.0      & 49.5  & 144    \\
H200 NVL            & 141    & HBM3e  & N5    & 814.0      & 49.5  & 211.5  \\
H200 SXM5           & 141    & HBM3e  & N5    & 814.0      & 49.5  & 211.5  \\
B200 SXM            & 192    & HBM3e  & N5    & 814.0      & 99.0  & 288    \\
Graphcore IPU GC200 & 64     & DDR4   & N7    & 823.0      & 47.1  & 96     \\
Habana Gaudi2       & 96     & HBM2e  & N7    & 826.0      & 47.3  & 144    \\
Ascend 910a         & 96     & HBM2e  & N7    & 456.0      & 22.2  & 144    \\
Graphcore BOW IPU   & 64     & DDR4   & N7    & 823.0      & 47.1  & 96     \\
\hline
\end{tabular}
}

\end{table*}

\begin{table*}[htbp]
\caption{Contextual information by location in 2025}
\label{tab:countries}
\centering
\resizebox{0.7\textwidth}{!}{%
\begin{tabular}{llll}
\hline
Location & Carbon Intensity & Electricity Price & Transport Footprint \\
         & (tCO$_2$-eq/kWh) & (\$/kWh)          & (kgCO$_2$eq/kg) \\
\hline
US            & 0.403     & 0.18    & 8.06    \\
Germany       & 0.334     & 0.36    & 5.61    \\
UK            & 0.175     & 0.37    & 6.16    \\
France        & 0.033     & 0.32    & 6.20    \\
Australia     & 0.489     & 0.27    & 6.76    \\
Netherlands   & 0.263     & 0.27    & 5.94    \\
Russia        & 0.367     & 0.06    & 4.49    \\
Italy         & 0.274     & 0.46    & 6.12    \\
Poland        & 0.703     & 0.24    & 5.33    \\
Spain         & 0.124     & 0.23    & 6.88    \\
China         & 0.514     & 0.08    & 0.00    \\
Brazil        & 0.091     & 0.15    & 12.77   \\
Switzerland   & 0.041     & 0.4     & 6.11    \\
Hong Kong     & 0.594     & 0.18    & 1.50    \\
World Average & 0.4       & 0.15    & 7.5
\end{tabular}
 }
\end{table*}

\begin{table*}[htbp]
\caption{Energy efficiency rate thresholds by countries for different upgrade cycle policies. Each cell represents the months to double energy efficiency.}
\label{tab:eeff_countries}
\centering
\resizebox{0.5\textwidth}{!}{%
\begin{tabular}{llll}
\hline
Location  & \multicolumn{3}{c}{Months to double energy efficiency} \\
          & \multicolumn{3}{c}{with different upgrade cycles} \\
          & 4 years & 5 years & 6 years \\
\hline
US            &     604     &     938    &   1827   \\
Germany       &     505     &     785    &   1526   \\
UK            &     272     &     420    &    812   \\
France        &      63     &      95    &    175   \\
Australia     &     731     &    1137    &   2216   \\
Netherlands   &     401     &     622    &   1207   \\
Russia        &     555     &     862    &   1678   \\
Italy         &     417     &     647    &   1256   \\
Poland        &    1046     &    1630    &   3183   \\
Spain         &     197     &     303    &    583   \\
China         &     778     &    1210    &   2360   \\
Brazil        &     147     &     225    &    431   \\
Switzerland   &      75     &     113    &    211   \\
Hong Kong     &     893     &    1391    &   2714   \\
\end{tabular}
 }
\end{table*}

Table~\ref{tab:trends_top_perf} presents the empirical trends in performance and energy efficiency improvements using the MLPerf benchmark data across device generations for the fastest executions per year. Table~\ref{tab:trends_top_eeff} shows the trends for the most energy-efficient executions per year. All means and standard deviations are computed in the logarithmic scale, where the regression is performed, then transformed into months to double performance/efficiency for interpretability. All benchmarks are normalized by the number of nodes. We show the MLPerf IDs selected per benchmark for data reproducibility.

\begin{table*}[ht]
\caption{Performance and energy efficiency improvement trends for AI platforms, based on the fastest executions per year from MLPerf Training~\cite{mattson2020mlperf} and MLPerf HPC ~\cite{MLPerf_HPC_2021} benchmarks.}%
\label{tab:trends_top_perf}
\resizebox{\textwidth}{!}{%
\begin{tabular}{@{}lllllll@{}}
\toprule
                  & \multicolumn{4}{c}{Months to Double (avg$\pm$std)}                &  Benchmark  &  IDs  \\
Benchmark         & \multicolumn{2}{c}{Performance} & \multicolumn{2}{c}{Energy Eff.} &  Database   &       \\
\midrule
bert              & 15 & $\pm$  1 &  21 & $\pm$  3 &  MLPerf Training    &  v0.7.40, v0.7.45, v1.0.1093, v2.0.2070, v2.1.2002,                              \\
                  & & & & & &  v2.1.2054, v4.0.20, v5.0.36                                                                                                  \\
dlrm              & 19 & $\pm$  7 &  31 & $\pm$ 17 &  MLPerf Training    &  v0.7.43, v2.0.2015, v2.0.2056, v2.1.2091                                        \\
dlrm\_dcnv2       & 16 & $\pm$  8 &  25 & $\pm$ 11 &  MLPerf Training    &  v4.0.38, v3.1.2080, v4.1.33, v5.0.75                                            \\
gpt3              & 11 & $\pm$  1 &  15 & $\pm$  3 &  MLPerf Training    &  v3.1.2033, v4.1.57, v4.1.82                                                     \\
llama2\_70b\_lora &  8 & $\pm$  0 &  10 & $\pm$  1 &  MLPerf Training    &  v4.0.40, v4.0.37, v4.1.68, v5.0.22                                              \\
llama31\_405b     & 10 & $\pm$  0 &  18 & $\pm$  0 &  MLPerf Training    &  v5.0.14, v5.0.66                                                                \\
mask\_r\_cnn      & 29 & $\pm$  9 &  46 & $\pm$ 17 &  MLPerf Training    &  v0.7.40, v0.7.45, v0.7.12, v2.1.2039, v3.0.2045,                                \\
                  & & & & & &  v3.1.2080, v3.0.2064                                                                                                         \\
nmt               & 31 & $\pm$ 11 &  66 & $\pm$ 12 &  MLPerf Training    &  v0.7.40, v0.7.45, v0.7.19                                                       \\
resnet            & 31 & $\pm$  3 &  69 & $\pm$ 16 &  MLPerf Training    &  v0.7.39, v0.7.1, v1.0.1096, v2.0.2034, v3.0.2005,                               \\
                  & & & & & &   v3.1.2037, v4.0.23, v4.0.71                                                                                                 \\
retinanet         & 34 & $\pm$ 22 & 102 & $\pm$ 70 &  MLPerf Training    &  v2.1.2013, v3.0.2004, v4.0.19, v4.0.23, v5.0.87                                 \\
rgat              & 10 & $\pm$  0 &  19 & $\pm$  0 &  MLPerf Training    &  v4.1.29, v5.0.74                                                                \\
ssd               & 43 & $\pm$ 29 &  69 & $\pm$ 50 &  MLPerf Training    &  v0.7.40, v0.7.45, v0.7.12, v1.1.2050, v1.1.2023                                 \\
stable\_diffusion &  6 & $\pm$  1 &   9 & $\pm$  2 &  MLPerf Training    &  v4.0.75, v4.1.33, v5.0.36                                                       \\
transformer       & 31 & $\pm$ 12 &  65 & $\pm$ 18 &  MLPerf Training    &  v0.7.40, v0.7.45, v0.7.19                                                       \\
unet3d            & 15 & $\pm$  7 &  58 & $\pm$ 24 &  MLPerf Training    &  v4.0.24, v4.0.23, v4.0.71                                                       \\
cosmoflow         & 13 & $\pm$  3 &  13 & $\pm$  3 &  MLPerf HPC TTT     &  v0.7.404, v0.7.413, v2.0.8006, v3.0.8006                                        \\
deepcam           & 31 & $\pm$ 10 &  33 & $\pm$  9 &  MLPerf HPC TTT     &  v1.0.1107, v1.0.1119, v1.0.1123, v3.0.8000, v3.0.8007                           \\
opencatalyst      & 15 & $\pm$  4 &  16 & $\pm$  4 &  MLPerf HPC TTT     &  v1.0.1108, v1.0.1118, v3.0.8003, v3.0.8008                                      \\
\midrule
Mean              & 15 & $\pm$  9 &  23 & $\pm$ 16  &  & \\
\bottomrule
\end{tabular}%
}
\end{table*}

\begin{table*}[hp]

\caption{Performance and energy efficiency improvement trends for AI platforms, based on the most energy efficient executions per year from MLPerf Training~\cite{mattson2020mlperf} and MLPerf HPC ~\cite{MLPerf_HPC_2021} benchmarks.}%
\label{tab:trends_top_eeff}
\resizebox{\textwidth}{!}{%
\begin{tabular}{@{}lllllll@{}}
\toprule
                  & \multicolumn{4}{c}{Months to Double (avg$\pm$std)}                &  Benchmark  &  IDs  \\
Benchmark         & \multicolumn{2}{c}{Performance} & \multicolumn{2}{c}{Energy Eff.} &  Database   &       \\
\midrule
bert              & 15 & $\pm$  1 &  21 & $\pm$  3 &  MLPerf Training    &  v0.7.40, v0.7.45, v1.0.1093, v1.0.1087, v2.1.2002,                              \\
                  & & & & & &  v2.1.2054, v4.0.20, v5.0.36                                                                                                  \\
dlrm              & 19 & $\pm$  7 &  31 & $\pm$ 17 &  MLPerf Training    &  v0.7.43, v2.0.2015, v2.0.2056, v2.1.2091                                        \\
dlrm\_dcnv2       & 18 & $\pm$ 10 &  24 & $\pm$ 10 &  MLPerf Training    &  v4.0.38, v3.1.2080, v4.1.33, v5.0.32                                            \\
gpt3              & 22 & $\pm$ 70 &  13 & $\pm$  7 &  MLPerf Training    &  v3.1.2033, v3.1.2004, v4.1.82                                                   \\
llama2\_70b\_lora &  8 & $\pm$  0 &  10 & $\pm$  1 &  MLPerf Training    &  v4.0.40, v4.0.37, v4.1.68, v5.0.22                                              \\
llama31\_405b     & 10 & $\pm$  0 &  18 & $\pm$  0 &  MLPerf Training    &  v5.0.14, v5.0.66                                                                \\
mask\_r\_cnn      & 33 & $\pm$  9 &  41 & $\pm$ 14 &  MLPerf Training    &  v0.7.40, v0.7.45, v0.7.12, v1.1.2022, v3.0.2045,                                \\
                  & & & & & & v3.1.2080, v3.0.2048                                                                                                          \\
nmt               & 32 & $\pm$ 11 &  25 & $\pm$  9 &  MLPerf Training    &  v0.7.40, v0.7.45, v0.7.68                                                       \\
resnet            & 34 & $\pm$  4 &  58 & $\pm$ 21 &  MLPerf Training    &  v0.7.39, v0.7.1, v1.0.1096, v0.7.68, v3.0.2005,                                 \\
                  & & & & & & v3.1.2037, v3.0.2048, v4.0.71                                                                                                 \\
retinanet         & 62 & $\pm$ 51 &  74 & $\pm$ 48 &  MLPerf Training    &  v2.1.2013, v3.0.2004, v4.0.19, v3.0.2048, v5.0.43                               \\
rgat              & 10 & $\pm$  0 &  19 & $\pm$  0 &  MLPerf Training    &  v4.1.29, v5.0.74                                                                \\
ssd               & 45 & $\pm$ 30 &  45 & $\pm$ 36 &  MLPerf Training    &  v0.7.40, v0.7.45, v0.7.12, v0.7.68, v1.1.2023                                   \\
stable\_diffusion &  6 & $\pm$  1 &   9 & $\pm$  2 &  MLPerf Training    &  v4.0.75, v4.1.33, v5.0.36                                                       \\
transformer       & 37 & $\pm$ 13 &  27 & $\pm$ 11 &  MLPerf Training    &  v0.7.40, v0.7.45, v0.7.68                                                       \\
unet3d            & 18 & $\pm$  1 &  45 & $\pm$ 47 &  MLPerf Training    &  v4.0.24, v4.0.17, v4.0.71                                                       \\
cosmoflow         & 13 & $\pm$  2 &  13 & $\pm$  3 &  MLPerf HPC TTT     &  v0.7.404, v0.7.413, v2.0.8002, v3.0.8006                                        \\
deepcam           & 31 & $\pm$ 10 &  33 & $\pm$  9 &  MLPerf HPC TTT     &  v1.0.1107, v1.0.1119, v1.0.1123, v3.0.8000, v3.0.8007                           \\
opencatalyst      & 16 & $\pm$  3 &  16 & $\pm$  4 &  MLPerf HPC TTT     &  v1.0.1108, v1.0.1118, v2.0.8002, v3.0.8008                                      \\
\midrule
Mean              & 16 & $\pm$  11 &  21 & $\pm$ 12  &  & \\
\bottomrule
\end{tabular}%
}
\end{table*}

\end{document}